\documentclass{ws-procs9x6}

\newcommand{\bea}{\begin{eqnarray}}
\newcommand{\eea}{\end{eqnarray}}
\newcommand{\bib}{\bibitem}

\begin{document}

\title{
Cosmology at  the Turn of Centuries 
}

\author{
A.D. Dolgov
}

\address{
INFN, sezione di Ferrara,
Via Paradiso, 12 - 44100 Ferrara,
Italy\\
and\\
ITEP, Bol. Cheremushkinskaya, 25, Moscow, Russia\\
E-mail: dolgov@fe.infn.it}

\maketitle

\abstracts{
A brief review of the present-day development of cosmology is presented for mixed physical 
audience. The universe history is briefly described. Unsolved problems are discussed, in particular, the mystery of the cosmological constant and dark energy, the problems of dark 
matter, and baryogenesis.  A brief discussion of the cosmological role of neutrinos is also
presented.
}

\section{Introduction \label{s-intr}}

Probably the most important scientific breakthrough of
the last quarter of the previous century was related to tremendous progress
in cosmology. From a poor relative of distinguished family of fundamental
science characterized by the words prescribed to L. Landau: ``always in
error but never in doubt'', cosmology turned, or is turning, into an exact 
science and possibly the most interesting one. It is full of mysteries, 
unsolved problems, puzzling phenomena and strongly indicates that there
exists new physics beyond the standard model. 

The stunning development of cosmology is due to a combination of two 
factors: 1) development of new theoretical methods, application of particle
physics and quantum field theory
to the early and contemporary universe; 2) new observational 
technique and devices, much more precise then even recently used ones and 
exploration of new windows to the universe (electromagnetic radiation in 
all wave length ranges, gravitational waves, neutrinos, high energy cosmic 
rays). Taken together they have led to Great Cosmological Revolution which 
keeps on going, turning into ``permanent revolution'' almost on Leo Trotsky 
own terms but, fortunately, not so bloody.

Cosmological parameters are now quite accurately known with even
brighter perspectives for the 
nearest future. One impressive example is the baryon 
asymmetry of the universe or, better to say, the ratio of the 
cosmological number 
density of baryons to the number density of photons in cosmic microwave
background radiation (CMBR). When the first works on the baryon asymmetry
were written this ratio was known as $\beta = n_B/n_\gamma = 10^{-9\pm1}$.
Now it is $\beta = (6\pm 1)\cdot 10^{-10}$. The progress is striking. 

\section{Cosmological Parameters \label{s-cosm-par}}

An important indication that cosmology is entering the club of exact 
sciences is the precision in determination of the values of basic
cosmological parameters. The progress of the recent years is achieved,
to a large extent, thanks to the measurements of the angular fluctuations 
of the CMBR\cite{cmbr} and detailed study of large scale structure (LSS)
of the universe\cite{lss}. 

The universe expansion law, $V=Hr$, is usually expressed in terms of the 
dimensionless Hubble parameter $h = H/100$ km/sec/Mpc. According to the 
present day 
measurements, it is $h=(0.72-0.73)\pm 0.05$. Previously for a long time it's 
value was between 1 and 0.5 with unclear systematic arrows.

With this accurate determination of $h$ the critical value of the cosmological
energy density becomes
\bea
\rho_c = \frac{3H^2 m_{Pl}^2}{8\pi} = 10^{-29}\,{\rm g/cm}^3\, (h/0.73)^2=
5.62\,{\rm kev/cm}^3\,(h/0.73)^2
\label{rho-c}
\eea
The contribution of any form of matter into total cosmological energy density
is expressed in terms of the parameter $\Omega_j = \rho_j/\rho_c$;
sometimes a more accurately determined quantity is $\Omega_j h^2$. 

The total energy density in the universe is quite close to the critical
one, $\Omega_{tot} = 1.02\pm 0.02$ in agreement with inflationary 
theory\cite{guth}. Visible matter (either shining or absorbing light)
gives a minor contribution into total mass, 
$\Omega_{vis} \approx 0.005$ (see e.g. ref.\cite{vis}). The total fraction
of baryonic matter is about an order of magnitude larger,
$\Omega_b = 0.044\pm 0.004$. The answer to the question where are the 
unseen 90\% of baryons is not completely clear by now. 
A much larger contribution to $\Omega$ comes from some unknown form of 
matter, the so called dark matter, while a better word could be 
``invisible'', $\Omega_{DM} = 0.27\pm 0.04$. Though unknown what, it is 
still normal matter, presumably, though not surely,
with a normal non-relativistic equation of state with zero pressure 
$p=0$ and positive energy (mass) density, $\rho>0$. Dark matter is 
believed to posses the usual gravitational interactions. 

Much more puzzling, even mysterious, is the dominant contribution to the
cosmological energy density, the so called dark energy, with 
$\Omega_{DE} = 0.73\pm 0.04$. It has negative pressure density,
\bea
p=w\rho
\label{p-of-rho}
\eea 
with $w<-0.8$, so it could be vacuum energy (or, in other words,
cosmological constant) for which $w=-1$. Its impact on the cosmological 
expansion is anti-gravitating, i.e. it leads to accelerated 
expansion\cite{sn}. This statement follows from the Friedman/Einstein
equation for the second time derivative of the cosmological scale factor
$a(t)$:
\bea
\frac{\ddot a}{a} = - \frac{4\pi G}{3}\,\left(\rho + 3 p\right)
\label{ddot-a}
\eea
and acceleration of the expansion would be positive if $p<-\rho/3$, of course 
under the standard assumption that $\rho$ is positive. For negative $\rho$ anti-gravity 
could be easily created but such theories would possess quite unpleasant pathological 
features if special care is not taken.

In addition to the direct observation of the accelerated expansion by
high red-shift supernovae, there is an indirect argument in favor of
non-zero vacuum-like energy. Namely with $\Omega = 1$ and $h=0.73$
the age of the universe would be about 9 Gyr, while nuclear-chronology and 
stellar evolution theory give the number 12-14 Gyr. With non-zero vacuum
energy, $\Omega_{vac} =0.7$ the universe needs considerably more time
to reach the present-day state and its age would well agree with the data.
Theory of large scale structure formation also supports a non-zero value of the
vacuum energy.

One more comment is worth making. Different forms of matter/energy density
have similar values at the present time: 
$\Omega_b =0.044$, $\Omega_m=0.27$, $\Omega_{DE}=0.73$, though  
their physical origin could be quite different and unrelated at least at the level of 
our present understanding. They may naturally differ by many orders of magnitude. 
Moreover, densities of non-relativistic matter and dark energy 
evolve in different ways in the course of the cosmological
expansion (see sec.~\ref{s-cosm-con}). This makes the problem even more
profound. At the moment this cosmic coincidence (or cosmic conspiracy)
problem is not understood at all.

\section{Cosmological Constant/Dark Energy \label{s-cosm-con}}

Cosmological constant (or lambda-term)
was introduced to equations of general relativity
by Einstein in 1918 when he found out that the equations did not admit 
stationary solution in cosmological situation with homogeneous
and isotropic matter source $T_{\mu\nu}$:
\bea
R_{\mu\nu} - \frac{1}{2}\, g_{\mu\nu} R - \Lambda g_{\mu\nu} =
8\pi G_N T_{\mu\nu}
\label{lambda}
\eea
Later, after the Friedman's cosmological solution and 
the Hubble's observation of the universe expansion,
Einstein strongly rejected the idea of cosmological constant and considered 
it as the ''biggest blunder'' of his life. For a long time after that the
majority of the astronomical/cosmological establishment even refused to
hear about it, though were were notable exceptions like e.g. Lemaitre, 
De Sitter, Eddington. The attitude of majority could be characterized by
the words written by G. Gamow in his autobiography book ``My World Line'':
``lambda rises its nasty head again'', after astronomical data indicated an
accumulations of quasars near red-shift $z=2$. In a sense Gamow was right 
because these data were explained without cosmological constant but it seems
impossible to avoid non-zero $\Lambda$ today.

According to the contemporary point of view, $\Lambda$-term can be 
interpreted as the energy-momentum tensor of vacuum and should be positioned
in the r.h.s. of the Einstein equation:
\bea
R_{\mu\nu} - \frac{1}{2}\, g_{\mu\nu} R  =
8\pi G_N \left( T_{\mu\nu}^{(m)} + \rho_{vac} g_{\mu\nu}\right)
\label{vacuum}
\eea
Quantum field theory immediately leads to a very serious trouble\cite{ybz}.
The energy of the ground state is generally non-vanishing and, moreover,
for a single species it is infinitely divergent:
\bea
\rho_{vac} = g_s\int \frac{d^3 p} {(2\pi)^3 }\sqrt{ p^2 + m^2} = \infty ^4
\label{rhob}
\eea 
Here $m$ is the mass of the field, $g_s$ is the number of spin states of 
the field and it is assumed that the field in question is a bosonic one. 
One cannot live in the world with infinitely big vacuum energy, so Zeldovich
assumed that bosonic vacuum energy should be compensated by vacuum
energy of fermionic fields. Indeed, vacuum energy of fermions 
is shifted down below zero and is given by exactly the same integral as
(\ref{rhob}) but with the opposite sign. (This is related to the condition that
bosons are quantized with commutators, while fermions are quantized with 
anti-commutators.) So, if there is a symmetry between bosons and fermions 
such that for each bosonic state there exists a fermionic state with the
same mass and vice versa, then the energy of vacuum fluctuations of bosons and
fermions would be exactly compensated, giving zero net result. This 
assertion \cite{ybz} was made before the the pioneering works on 
super-symmetry \cite{gl} were published. However, since super-symmetry is 
broken this compensation is not complete and non-compensated amount of 
vacuum energy in any softly-broken super-symmetry must be non-vanishing,
$\rho_{vac}^{(susy)} \sim m^4_{susy} \geq 10^8$ GeV$^4$. This is 55 orders
of magnitude larger than the cosmological energy density 
$\rho_c \approx 10^{-47}$ GeV$^4$, eq.~(\ref{rho-c}). 
Supergravity, i.e. locally
realized supersymmetry, allows vanishing vacuum energy even in the broken 
phase but at the expense of fantastic fine-tuning by more than 120 orders
of magnitude. Somewhat smaller contribution but still significant (mismatch
by 55 orders of magnitude), and in a sense more troubling, comes from the 
vacuum condensates in quantum chromodynamics (QCD). It is practically an
experimental fact that vacuum in QCD is not empty but filled by 
quark\cite{gm} and gluon condensates\cite{svz} with the energy density
about $10^{-2}-10^{-4}$ GeV$^4$. What else ``lives'' in vacuum and exactly
``eats up'' all these contributions (with accuracy $10^{-45}$ or maybe even 
$10^{-120}$) is the biggest mystery in modern physics.
The reviews of vacuum energy problem and possible solutions (none is
satisfactory at the moment) can be found in refs.\cite{lam-rev}.

Another side of the problem emerges from the observed acceleration of the
universe: it means that either the vacuum energy-momentum tensor is 
non-vanishing, $T_{\mu\nu} = \rho_{vac} g_{\mu\nu} \neq 0$,
or there exists something unknown with negative and large 
by absolute value pressure, $p=w\rho$ with $w<-1/3$.
According to the covariant conservation of energy-momentum:
\bea 
\dot \rho = -3H (\rho +p),
\label{dot-rho}
\eea
the energy density of matter with the equation of state~(\ref{p-of-rho})
evolves in the course of expansion as $\rho\sim a^{-3(1-w)}$. 
In particular, for vacuum energy $\rho = const$. A puzzling feature
is why the energy densities of non-relativistic matter which evolves
as $\rho_m \sim a^{-3}$ and dark (vacuum) 
energy are almost coinciding today, though their cosmological evolution 
were completely different. This may be explained by the so called 
quintessence\cite{qnt}, that is by a new scalar field with 
vanishingly small mass and negligible interaction with other matter
fields, except for gravity. Equation of motion of this field might have
a tracking solution whose energy density closely follows the dominant 
one, though a fine-tuning is necessary to realize such a solution.
On the other hand, even if the problem of coincidence of $\rho_{vac}$
and $\rho_m$ today, may be solved by the tracking solution,
this new field does not explain why the vacuum energy is so close to zero.

A model which in principle could solve both these problems is based on
dynamical adjustment of vacuum energy by a new massless scalar\cite{scal},
vector\cite{vec}, or tensor\cite{ten} field coupled to gravity in such
a way that this new field is unstable in De Sitter space-time with 
respect to formation of vacuum condensate whose energy would compensate
the source (i.e. the original vacuum energy) in accordance with the 
general principle of Le Chatellier. Such models have a generic property
that the vacuum energy is compensated, however the compensation is not 
complete but only down to the terms of the order of the critical energy 
density at the running time moment $t$. So at each moment there is a 
non-compensated remnants (dark energy) with $\Omega \sim 1$ with equation
of state which may differ very much from the usual ones. Phenomenologically
such models describe decaying vacuum energy and contain many of the later
suggested ``quintessentuial'' ideas. The only, but important, shortcoming
is that a realistic cosmological model based on this mechanism has not
yet been found.

An important quantity which may lit some light on the physical nature of
dark energy is the value of the parameter $w$ which describes the
equation of state of dark energy~(\ref{p-of-rho}). The present-day data 
agrees with $w=-1$, i.e. dark energy could be
the vacuum energy. However, other
values of $w$ are not excluded, moreover, it is possible that $w$ is not a
constant but a function of time and even that equation of state does not
exist, so $p$ cannot be expressed locally through $\rho$.

In a normal field theory the energy dominance condition, $|p|<\rho$
is usually fulfilled. However, dealing with such a strange entity as the
dark energy we cannot exclude that $w<-1$. In this quite unusual case the
energy density rises(!) in the course of expansion, $\dot \rho >0$ (see 
eq.~(\ref{dot-rho})). An example of field theory leading to such a regime
can be found in ref.\cite{ten}: it is a theory of free massless vector
(or tensor) field with the Lagrangian density 
${L} = V_{\mu;\nu}V^{\mu;\nu}/2 $. Such a theory in a curved background
is unstable with 
respect to formation of vacuum condensate of the time component $V_t$
with the energy and pressure densities:
\bea
\rho_V &=& \dot V_t^2/2 + 3H^2 V_t^2/2, 
\label{rho-v}\\ 
p_V &=& \dot V_t^2/2 - 3H^2 V_t^2/2 -\dot HV_t^2 -2H\dot V_t V_t 
\label{p-v}
\eea
There is no equation of state $p=p(\rho)$ in this theory but anyhow
$\rho_V + p_V $ may be negative and a new type of cosmological singularity:
\bea
 H\sim (t_0 -t)^{-3/2}
 \label{h-sing}
 \eea
  can be reached in a final time resulting in 
``tearing apart'' the universe\cite{ten}. A different model of negative
$(\rho+p)$ based on a scalar field with a wrong sign of the kinetic term
and a discussion of a similar singularity can be found in 
ref.\cite{phantom}. The pathological properties of such theories
probably indicate that a negative $\rho + p$ is impossible, though it 
is not yet rigorously forbidden.

Understanding the problem of vacuum energy compensation may have
a noticeable impact on cosmology (and quite probably on fundamentals of
quantum field theory). The mechanism that ensures this compensation may
change the standard picture of the cosmological evolution. However,
at the moment there is no strong demands for such changes. On the other hand,
some discrepancies which may exist in big bang nucleosynthesis (see below
sec.~\ref{ss-bbn}) or possibly in the formation of large scale structure 
could be cured by the dark energy. A related strong 
challenge is to understand what is the dark energy and
a very important information about it can be obtained through a
more accurate measuring of $w$ in the future.

\section{Dark Matter \label{s-dm}}

Second most important unsolved problem in cosmology (which may also have
a strong impact on particle physics and field theory) is the
problem of dark matter. The latter, most probably, consists from normal 
but yet undiscovered particles or fields. There are many (maybe, too many) 
possibilities discussed in the literature and at the moment we do not know
which one of these hypothetical objects plays the role of dominant matter
constituent of our world, or maybe there are even several of them. For a 
recent review see e.g. the lectures\cite{olive-03}

Though the dark matter is not observed directly its existence seems to be
firmly established. First, the total fraction of non-relativistic matter in 
the universe is $\sim 0.3$, while the amount of baryonic contribution is 
about 0.05 (see sec.~\ref{s-cosm-par}). The latter is determined
by two independent methods: by measurements of angular fluctuations of
CMBR and by observations of abundances of light elements produced at BBN. 

Another argument in favor of cosmological domination of non-baryonic 
matter is based on the
theory of large scale structure formation. Structures in baryo-dominated
universe can only be formed after hydrogen recombination which took place
at $T\approx 3000$ K, i.e. at red-shift $z_{rec}=10^3$ when the matter became 
electrically neutral. Before that period a large light pressure experienced by 
electrons (and as a result, by protons) prevented them from gravitational
clusterization. After recombination, initially small density 
perturbations, $\delta\rho/\rho$ rose
as the cosmological scale factor and hence could increase only by the factor
$z_{rec}=10^3$. On the other hand, the measured angular fluctuations
of CMBR temperature are quite small, $\delta T/T<$(a few)$\times 10^{-5}$.
Hence it  follows under the standard assumption of adiabatic density perturbations 
(this assumption is now confirmed by CMBR data) that the latter should be also
small at recombination, $\delta\rho/\rho< 10^{-4}$. Hence they should
remain small at the present epoch, even after amplification by 
3 orders of magnitude.
So one could conclude that there should be some other form of matter
which does not interact with light and which started to form structures 
long before recombination. On the other hand, much larger density 
perturbations at small scales are not formally excluded by observations
and they might allow an efficient structure formation in purely baryonic 
universe. Still, the combined data are very much against structure formation 
without dark matter. In particular, inflation predicts flat spectrum of 
perturbations with spectral index $n=1$. From observations follows
$n=0.93\pm 0.03$. If this spectral behavior remains true at small scales
then larger density perturbations, mentioned above, would not be present. 

Most probably cosmological dark matter consists of cold relics from Big
Bang. It is the so called cold dark matter (CDM). Two most popular candidates
for the latter are the lightest supersymmetric particle (which could be as 
heavy as a few hundreds GeV) and axions (which would be extremely light,
about $10^{-5}$ eV). However, such simple forms of dark matter meet serious
problems in description of details of large scale structure. There are 
several troubling features. In particular, CDM cosmology predicts dark 
halos with steep cusps\cite{cusp}, while observations indicate that 
halos have a constant density cores (see e.g. ref.\cite{blok-03}). Theory
also predicts too many, by factor 5, galactic satellites\cite{stlt}.
Comparison\cite{cluster} 
of galactic cluster abundances at high ($z>0.5$) and low redshifts
is compatible with the theory of cluster evolution for a very low 
cosmological matter density, $\Omega_m \approx 0.17$,
which is much smaller than the value, $\sim 0.3$, obtained by other methods.
And at last, the galaxies in CDM simulations have considerably smaller
angular momenta, than those observed (see the paper\cite{angl-mom} and 
references therein). All these inconsistencies could be either prescribed to 
shortcomings of numerical simulations and, in particular, to neglecting 
some essential physical processes, or, more probably, to a real crisis
in cold dark matter cosmology. 

Several new forms of dark matter were suggested to overcome the CDM 
problems: warm dark matter (WDM), self-interacting dark matter 
and, in particular, long discussed mirror or shadow matter (which is a 
special case of self-interacting dark matter), decaying cold dark matter, etc.
Models with non-canonical forms of the spectrum of primordial fluctuations
and models with mixed adiabatic and isocurvature fluctuations were also 
considered. (For a recent review and a list of references 
see the papers\cite{dm-rev}, here is neither space not time to discuss these issues
in detail.) It seems that a simple canonical model with one form of dark
matter and flat (inflationary) spectrum of perturbations does not 
satisfactory describe details of cosmic large scale structure and some 
deviations from the standard scenario seem to be necessary. It makes the
situation more complicated and more interesting. It is worth mentioning 
that models with several forms of dark matter, e.g. mixed CDM+WDM, make the
problem of cosmic conspiracy even more profound because, in addition to 
similar magnitudes of $\Omega_b$, $\Omega_m$, and $\Omega_{DE}$
one has to explain close contributions to $\Omega$ of several new 
forms of DM.

\section{Main events in cosmological evolution \label{s-events}}

Here we will briefly discuss the history of the universe and 
physical phenomena necessary for creation of the present state
of the world where we can live.

\subsection{Before beginning \label{ss-before}}

Nothing is known about the universe prior for a certain temporal moment.
So we cannot extend our history infinitely backward in time. It may be so
because theory of quantum gravity is yet missing and there is no way to
describe the state of the universe when the characteristic curvature
and energy density had Planckian magnitudes. Possibly even time and space
in our classical understanding did not exist ``at that time''. There are
some attempts to go beyond Planck scales or, better to say, before
Big Bang exploring e.g. string cosmology\cite{string-cosmo}. However, it 
is difficult to say if this or any other continuation through big-bang was 
indeed realized. Another possibility of periodically oscillating universe
between big bang and big crunch was suggested ages ago (a list of the
early papers and discussion can be found in the books\cite{osc-univ})
and was recently revitalized\cite{epokr} as an alternative to inflation.
Such models have an evident difficulty: in contraction phase already evolved
large inhomogeneities become even larger and catastrophic formation of
black holes during contraction phase may endanger the scenario. 
A recent discussion of behavior of perturbations in oscillating or
bouncing universe can be found in the papers\cite{pert-osc}.

\subsection{Inflation \label{ss-infl}}

The earliest period of the universe evolution whose existence seems to
be on firm grounds is the inflationary epoch. Today one could even say 
that inflation is an experimental fact. After the original suggestion by 
Guth\cite{guth} and first realistic models\cite{infl-mod} there appeared
zillions of papers on the subject; for recent reviews one can address
ref.\cite{infl-rev}

During relatively short inflationary period, when the universes exponentially
expanded, $a(t)\sim \exp (H_I t)$, the proper initial conditions for 
creation of the present-day universe have been secured. During this period
the cosmological energy density was, by assumption, dominated by a slow
varying scalar field, inflaton, with the vacuum-like energy-momentum
tensor, $T_{\mu\nu}\approx g_{\mu\nu} \rho_I$. As we have already mentioned
above, such energy-momentum tensor creates cosmological gravitational
repulsion and hence it explains the origin of the observed today expansion.
If inflation was sufficiently long, $H_I \Delta t_I \geq 70$, then the 
universe would be flat, $\Omega = 1\pm 10^{-4}$ in accordance with 
observations (in fact the necessary duration of inflation depends upon the
temperature of the universe heating and might be somewhat smaller than 70
Hubble times). The universe would be almost homogeneous and isotropic on 
large scales. This explains why CMBR coming from different directions has
almost the same temperature - one should keep in mind that without inflation
the regions on the sky separated by more than one degree would never 
communicate to each other. Simultaneously with the ``smoothing down'' the
universe, inflaton created small density perturbations but at 
astronomically large scales which much later became seeds of large scale
structure formation. In non-inflationary cosmology no reasonable mechanism 
of creation of density perturbations was known and the problem of their
generation at astronomically large scale remained a great mystery.

A unique explanation of these previously unexplainable features would be
enough to consider inflation as an established fact. Moreover, there are
some quantitative predictions of inflationary scenario supported by the
astronomical data. First is of course a prediction of flat universe, 
$\Omega =1$. Second  is a prediction of flat (Harrison-Zeldovich\cite{hz}) 
spectrum of perturbations with spectral index $n=1$. 
In fact inflationary spectrum usually slightly deviates from the flat 
one, depending upon the concrete model of inflation (for a review see
e.g. ref.\cite{pert-infl}). According to the recent WMAP
data (fifth paper in ref.\cite{cmbr}), the index deviates from unity,
$n=0.93\pm 0.03$. It could be a worrying sign but maybe the deviation is
not significant taking into account possible statistical and systematical
errors. On the other hand, as argued in ref.~\cite{mukhanov-03}, the graceful exit 
from inflation demands $0.92<n<0.97$ and the WMAP data can be considered
as confirmation of simple inflationary scenarios. However,
even if the spectral index happened to be outside these bounds, it
would not mean that inflation is
excluded. There could be e.g. some other forms of perturbations, in 
particular, isocurvature ones (see e.g. the review\cite{ad-bs} for 
possible mechanisms and recent papers\cite{iso-curv} and references
therein) which would have a different spectrum and which may 
explain a noticeable deviation of $n$ from unity.

So as a whole inflation is a great success. It is observationally 
confirmed and theoretically beautiful.
Now we have to fix some details: what is the inflaton, in which potential
it evolves, are there gravitational waves generated at inflationary stage,
etc. This is of course highly non-trivial, especially because
there could be some other sources of perturbations, except for the inflaton.
The measurements of CMBR angular spectrum and polarization gives some hopes
to progress in this direction.
Existence of inflation means in particular that there should be new physics
beyond the standard model (SM) because there is no space for the inflaton 
in the frameworks of SM.

\subsection{Baryogenesis \label{ss-bs}}

Predominance of matter over antimatter was one of the biggest cosmological 
puzzles of the first two thirds of the XX century. Its solution was outlined
by Sakharov\cite{sakharov-67} in 1967
and is now commonly accepted. The mechanism 
is based on three conditions:
1) Non-conservation of baryonic charge;
2) Breaking of C and CP symmetries;
3) Deviation from thermal equilibrium in primeval plasma.
All these three conditions are known to be true either from experiment 
or because they are well justified theoretically. Moreover, the existence
of the charge asymmetric universe itself is a strong indication to 
non-conservation of baryons, otherwise inflation would not be possible. 

The time and temperature interval where baryogenesis took place strongly
depends on concrete model and may vary in a very wide range from GUT or
even almost Planck scales down to a fraction of
GeV. For the reviews and more recent 
quotations see refs.\cite{ad-bs,bs-rev}. There are many models of 
baryogenesis considered in the literature. Possibly an incomplete list 
includes:
\begin{enumerate}
\item{}
Baryogenesis in out-of-equilibrium heavy particle decays\cite{sakharov-67}.
\item{}
Electroweak baryogenesis\cite{krs}.
\item{}
Baryogenesis by supersymmetric baryonic charge condensate\cite{susy-bs}.
\item{}
Spontaneous baryogenesis\cite{spont-bs}.
\item{}
Baryo-through-lepto-genesis\cite{lepto}.
\item{}
Baryogenesis in black hole evaporation\cite{bh-bs}.
\item{}
Baryogenesis by topological defects (domain walls, cosmic strings, 
magnetic monopoles)\cite{top-bs}.
\end{enumerate}
The problem is to understand what of the above is indeed realized and how
it can be verified. The second part is non-trivial because usually a model
of baryogenesis has to explain only one number the ratio of baryonic charge 
number density to the number density of photons in CMBR, 
$\beta =6\cdot 10^{-10}$, and there could be many models giving the same
baryon asymmetry. It seems established that electro-weak baryogenesis,
which might proceed in the frameworks of the standard model, is not efficient
enough to produce the observed asymmetry. All other models demand new 
physics. So the baryon asymmetry of the universe is another cosmological
fact that indicates to existence of physics beyond the standard model.

There are plenty of models of baryogenesis which simultaneously with excess of
matter over antimatter in our neighborhood predict that there could be
astronomically large domains of antimatter and maybe even not too far from
us. Such models are discussed in ref.\cite{anti-blois}. In this connection
expected BESS, Pamela, or AMS experiments searching for 
cosmic antinuclei $^4$He or heavier ones would be of primary importance
(for a review see e.g. ref.\cite{ams}).

\subsection{ Big Bang Nucleosynthesis \label{ss-bbn}}
 
Big bang nucleosynthesis (BBN) took place in a relatively old and cold
universe in the temperature interval from a few MeV down to 60-70 keV and
time in the range from roughly 1 sec up to 300 sec. Physical processes 
involved are well known: they include low energy weak interactions and 
low energy nuclear physics. In contrast to phenomena considered in the
previous subsections when we were in {\it terra incognita} and had to use 
our imagination based on reasonable theoretical models, now we are on firm
grounds of well established physics. 

The only unknown parameter that enters theoretical calculations of the
light element abundances is the ratio of baryon-to-photon number densities, 
$\beta= n_B/n_\gamma$, which a few years ago was found from BBN itself. 
Now can be independently determined from CMBR. A reasonable concordance 
of the two values of $\beta$
shows that the theory is in a good shape and that there was no influx of
photons into the cosmic plasma between neutrino decoupling at 
$T\approx 1$ MeV and hydrogen recombination at red-shift $z\approx 10^3$
or $T\approx 0.26$ eV. 
The accurate Bose-Einstein shape of the CMBR spectrum indicates the same
but in a slightly shorter temperature interval, roughly speaking, starting
from the red-shift $10^7$ (see e.g. the reviews\cite{dz,ad-nu}).

During those several minutes the following light elements have been 
produced: deuterium ($\sim  3\cdot 10^{-5}$, relative to hydrogen by 
number), helium-3 (about the
same number as $^2$H), helium-4 (23-24\% by mass), and a little of
$^7$Li (a few $\times \,\, 10^{-10}$). Accuracy of calculations is 
mostly determined by the uncertainties in the nuclear interaction rates
and is quite good. According to the analysis of ref.\cite{bbn-acc},
the theoretical accuracy is at the level $\leq 0.1\%$ for $^4He$, better 
than 10\% for $^2H$ and is about 20-30\% for $^7Li$. Anyhow, in all the 
cases theoretical uncertainty is much smaller than the observational 
precision. The latter suffers from two serious problems: systematic errors 
and poorly understood evolutionary effects. They are reviewed e.g. in
refs.\cite{ad-taup}. 
Despite existing uncertainties, theory reasonably well agrees with
observations. However, there are indications to possible discrepancies.
The results of different groups measuring deuterium at high redshifts,
which may be a primordial one, are in noticeable disagreement with each other.
Moreover, the measured abundances of $^4$He and $^2$H seem to correspond to
somewhat different values of $\beta$. It is not clear if these discrepancies are
serious and indicate some unusual physics (degeneracy of cosmic neutrinos,
non-negligible role of dark energy at BBN, some new particles present at BBN,
etc) or after a while all measurements will come to an agreement between 
themselves and with the value of $\beta$ inferred from CMBR. Hopefully it will
not take more than a decade.

\subsection{Large scale structure formation \label{ss-lss}}

During an initial period of the universe life-time (but after inflation)
the universe was very smooth, practically homogeneous and isotropic. It
remained such between the epoch of inflation and the end of radiation
domination (RD) era. This period lasted roughly speaking about 100,000 years
or until red-shift $z_{eq}\approx 10^4$. Small density perturbations
generated at inflation were almost frozen during RD period and were
practically unnoticeable. However, their importance at matter dominated
(MD) regime is difficult to overestimate. At MD-stage these small density
perturbations became unstable with respect to gravitational attraction,
they started to rise forming seeds from which astronomical large scale
structures such as galaxies, their clusters and superclusters evolved.
Development of such huge objects, their temporary evolution, and power
spectrum primarily depend upon the form of primordial density perturbations
and properties of dark matter. In particular, detailed study of the
large scale structure at different scales\cite{lss} can help to 
establish essential properties of dark matter particles prior to their  
possible discovery in direct experiments. To some extent this subject is 
discussed in sec.~\ref{s-dm}.

An interpretation of the astronomical data and conclusion about properties 
of DM-particles depend upon the hypothesis about the perturbation spectrum.
Usually the spectrum of primordial density perturbations is taken in one
parameter power law form with an arbitrary spectral index $n$. Such 
shape is justified
for the flat spectrum (with $n=1$) which is scale free and on dimensional 
grounds this is the only possible form of the spectrum. In the case that 
a dimensional parameter exists, any function of this parameter, and not
only a power law, is a priori allowed. Though, as is mentioned above, 
the nearly flat spectrum is supported by inflation and quite well agrees 
with the data, noticeable deviations from this type of spectrum are 
possible. Such deviations would be very interesting for physics of the
early universe and, in particular, for possible mechanisms of creation
of density fluctuations but simultaneously it would make it more
difficult to obtain conclusions about DM-particles from the LSS data.

\subsection{Future of the universe \label{ss-future}}

In Friedman cosmology with the normal matter content the ultimate 
fate of the universe is completely determined by the spatial curvature:
for flat (zero curvature) and open (negative curvature) geometry the 
universe will expand forever, while closed universe (positive curvature)
will stop expanding and will recollapse to big crunch. One can see that 
from the Friedman equation:
\bea
H^2 \equiv \left( \frac{\dot a}{a}\right)^2 =
\frac{8\pi \rho }{3 m_{Pl}^2} - \frac{k}{a^2}
\label{h-of-k}
\eea
where the sign of the constant $k$ determines the sign of curvature. 
The energy density $\rho$ of any normal matter with positive pressure
drops in the course of
expansion faster than $1/a^3$, the limiting case corresponds to 
non-relativistic matter with $p=0$. Hence the curvature
term will dominate at large $a(t)$ and  determine the universe destiny.

If the parameter $w$ connecting pressure and energy densities, 
eq.~(\ref{p-of-rho}), may be negative, then for $w<-1/3$
the pressure density which evolves
as $\rho\sim a^{-3(1+w)}$, would decrease slower than $k/a^2$ so
initially inessential curvature term would always remain such and
expansion would never stop in any geometry. Thus in a distant future
only gravitationally binded objects will continue to exist, while
distant galaxies will disappear from the sky. It would not create a big difference
for a naked eye if stellar luminosity remains the same during some billion years.
Unfortunately this is not so because the Sun and other stars will exhaust their 
nuclear fuel and the world will fall into (almost) complete darkness and cold which 
may be slightly lit up and heated by possible proton decay. For the proton life-time
$\tau_p \sim 10^{33}$ years the solar luminosity generated by the decay will
be 20 orders of magnitude smaller than the present-day solar luminosity.  
In this epoch the Earth
will be much stronger "illuminated" and heated by her own decaying protons. Nevertheless,
life will hardly be possible in such uncomfortable  conditions. The only chance for
survival could be catalysis of proton decay by e.g. magnetic monopoles\cite{rub-mon}
if they exist.  The consumption of the whole solar mass will allow to maintain life on the
Earth for about $10^{20}$ years. Using to this end other stars in the galaxy may extend
the life-time up to $10^{30}-10^{32}$ years, if other civilizations do not interfere.

All the story may end much faster if $p<-\rho$ and the universe will evolve to catastrophic
expansion singularity (\ref{h-sing}) in 10-100 Gyr. In this case not only astronomically large
objects but even atoms and elementary particles may be destroyed. However this
conclusion is model dependent and an inhomogeneous component of the dark energy
field $V_t(t,{\bf x})$ may stop the catastrophe.

\section{Neutrinos in cosmology \label{s-nus}}

The universe is filled with neutrinos - 55 neutrinos and the same number of antineutrinos
per cm$^3$ for each neutrino flavor - and though they are very light and weakly interacting 
their sheer number makes them cosmologically important. Knowing the number density of 
cosmic neutrinos one can immediately deduce an upper limit on their mass\cite{gz}:
\bea
\sum_a m_{\nu_a} < 94\,{\rm eV}\, \Omega_m h^2 = 15\,{\rm eV}
\label{gz-mnu}
\eea
where $\Omega_m = 0.3$ and $h^2 = 0.5$ have been substituted and the sum is taken over 
all neutrino species, $a=e,\mu,\tau$.

Since neutrinos were relativistic during a large part of the universe history, their
long free streaming path would suppress structure formation in neutrino dominated
universe at the scales below $M= 10^{17} M_\odot (m_\nu /{\rm eV})^{-2}$ where
$M_\odot$ is the solar mass. Hence massive neutrinos cannot be the dominant dark
matter particles and their mass density should be below $\Omega_\nu <  0.1$. Correspondingly
$\sum_a m_{\nu_a} < 5 {\rm eV}$ (further discussion and references to this and other subjects
discussed below can be found in the review\cite{ad-nu}). More detailed studies of the
large scale structure together with WMAP data on angular fluctuations of CMBR allowed
to improve this limit down to $\sum_a m_{\nu_a} < 0.7 {\rm eV}$. Already today astronomy is able
to constraint $m_\nu$ with better accuracy then direct experiment. Future Planck mission and
more data on LSS will possibly push this limit down to $\sim 0.1$ eV or will measure the
neutrino mass. This would be a unique example when the mass of an elementary particle
is determined by telescopes.

Apart from LSS, neutrinos can be traced through BBN and CMBR. BBN permits to constrain
the number of neutrino families\cite{n-nu}.  Keeping in mind the existing accuracy in extracting
primordial abundances of $^2$H and $^4$He from observations (see sec.~\ref{ss-bbn}) one can
impose the upper limit on the number of additional neutrino families, $\Delta N_\nu < 0.3-0.5$
with justified expectations to strengthen this limit down to about 0.1 in the near future. At the
present time the CMBR data are not competitive with the upper limit obtained from BBN. Still
the analysis\cite{n-nu-cmbr} of the CMBR data indicates, independently from BBN, that cosmological relic neutrinos indeed exist and the number of families is confined between: $1<N_\nu<6$. One may hope for a drastic improvement of this result
if the expected  sensitivity of the Planck mission at per cent level is achieved, so that  it may
register even non-equilibrium corrections to the energy spectrum of relic neutrinos at the
level of 3\%. These corrections are predicted to result from non-equilibrium 
$e^+e^-$- annihilation in the primeval plasma and deviations from ideal gas approximation
at the epoch when the universe was about 1 sec old (see discussion and references in the
review\cite{ad-nu}).

Consideration of BBN permits also to derive bounds on mixing between active and possible
sterile neutrinos, on the magnitude of cosmological charge asymmetry, on magnetic moment
of neutrinos, etc. 

\section{Conclusion \label{s-concl}}

Great progress in cosmology at the end of the previous century helped to understand the
universe much better and simultaneously led to discovery of many new puzzles, problems,
and even mysteries. Today cosmology tells us that fundamental physics is not completed
and there surely exist new phenomena outside well established known physics. We need to
discover much more to understand the observed features of the universe.

To start, the greatest mystery of (almost) complete cancellation of vacuum energy is not
(or cannot be) resolved in the frameworks of known physics. The next, possibly not so
striking, question about the nature of dark energy also remains unanswered. 

As for dark matter, it may possibly be explained by some extension of the realm of the
existing particles by addition either stable lightest supersymmetric particle or axion. There
are some more good candidates for dark matter particles or objects but it is not yet 
established which one of them makes dark matter.  Moreover, the CDM crisis (see 
sec.~\ref{s-dm}) possibly demand either several forms of DM, thus deepening the cosmic
conspiracy puzzle, or particles with rather strange properties unexpected in simple, 
theoretically motivated  generalizations of the standard model. 

Baryogenesis might in principle operate based on the existing minimal standard model of
particle physics but the latter is not sufficiently productive to create the observed baryon 
asymmetry of the universe. Hence new particles or fields are necessary.

Hopefully joint efforts of experimentalists and theoreticians will help to resolve some or,
in further perspective, all these problems but new mysteries and discoveries 
are waiting for us
on the way and this is what makes cosmology so interesting now. To conclude,
cosmology had very productive period during last quarter of the previous century, it is in
the process of exciting development today, and  bright future with many new discoveries
is coming.

\section{Acknowledgment}
The hospitality of the Research Center for the Early Universe of the University of Tokyo, where 
the preparation of this contribution for publication was completed, is gratefully acknowledged.

\end{document}